\documentclass[twocolumn]{aastex62}
\usepackage{natbib}
\usepackage{color}
\usepackage{mathptmx}
\usepackage{amsmath}
\usepackage{url}
\usepackage{multirow}
\usepackage{verbatim}

\graphicspath{{./}{figures/}}

\shorttitle{{\it AGORA} Comparison. Public data release}
\shortauthors{{\it AGORA} Collaboration et al.}

\begin{document}

\title{THE {\it AGORA} HIGH-RESOLUTION GALAXY SIMULATIONS COMPARISON PROJECT:  PUBLIC DATA RELEASE}

\author[0000-0002-6299-152X]{Santi Roca-F\`{a}brega$^{*}$}
\affil{Dept. F\`{i}sica de la Tierra y Astrof\`{i}sica, Facultad de Ciencias F\`{i}sicas, Instituto de F\`{i}sica de Part\`{i}culas y del Cosmos, Universidad Complutense de Madrid, E-28040 Madrid, Spain}
\affil{Center for Astrophysics and Planetary Science, Racah Institute of Physics, The Hebrew University, Jerusalem, 91904, Israel}
\email{$^*$Corresponding author:  sroca01@ucm.es}
\author[0000-0003-4464-1160]{Ji-hoon Kim}
\affil{Center for Theoretical Physics, Department of Physics and Astronomy, Seoul National University, Seoul, 08826, Korea}
\author[0000-0001-5091-5098]{Joel R. Primack}
\affil{Department of Physics, University of California, Santa Cruz, CA 95064, USA}
\author{Michael J. Butler$\textsuperscript{\textdagger}$}
\affil{Max-Planck-Institut f\"{u}r Astronomie, D-69117 Heidelberg, Germany}
\author{Daniel Ceverino$\textsuperscript{\textdagger}$}
\affil{Departamento de F\`{i}sica Te\`{o}rica, Universidad Aut\`{o}noma de Madrid, E-28049 Madrid, Spain}
\affil{Niels Bohr Institute, University of Copenhagen, 2100 Copenhagen Ø, Denmark}
\affil{Zentrum f\"{u}r Astronomie der Universit\"{a}t Heidelberg, D-69120 Heidelberg, Germany}
\author{Jun-Hwan Choi$\textsuperscript{\textdagger}$}
\affil{Department of Astronomy, University of Texas, Austin, TX 78712 , USA}
\author{Robert Feldmann$\textsuperscript{\textdagger}$}
\affil{Institute for Computational Science, University of Zurich, CH-8057 Zurich, Switzerland}
\author{Ben W. Keller$\textsuperscript{\textdagger}$}
\affil{Zentrum f\"{u}r Astronomie der Universit\"{a}t Heidelberg, D-69120 Heidelberg, Germany}
\author{Alessandro Lupi$\textsuperscript{\textdagger}$}
\affil{Scuola Normale Superiore, I-56126 Pisa, Italy}
\author{Kentaro Nagamine$\textsuperscript{\textdagger}$}
\affil{Theoretical Astrophysics, Department of Earth and Space Science, Graduate School of Science, Osaka University, Toyonaka, Osaka, 560-0043, Japan}
\affil{Kavli-IPMU (WPI), University of Tokyo, Kashiwa, Chiba, 277-8583, Japan}
\affil{Department of Physics and Astronomy, University of Nevada, Las Vegas, NV 89154, USA}
\author{Thomas R. Quinn$\textsuperscript{\textdagger}$}
\affil{Department of Astronomy, University of Washington, Seattle, WA 98195, USA}
\author{Yves Revaz$\textsuperscript{\textdagger}$}
\affil{Institute of Physics, Laboratoire d'Astrophysique, \'{E}cole Polytechnique F\'{e}d\'{e}rale de Lausanne, CH-1015 Lausanne, Switzerland}
\author{Romain Teyssier$\textsuperscript{\textdagger}$}
\affil{Institute for Computational Science, University of Zurich, CH-8057 Zurich, Switzerland}
\author{Spencer C. Wallace$\textsuperscript{\textdagger}$$^{15}$ for the {\it AGORA} Collaboration}
\affil{The authors marked with \textsuperscript{\textdagger}, in alphabetical order, contributed to the article by leading the effort within each code group to produce the simulation data.}





\begin{abstract}
As part of the {\it AGORA} High-resolution Galaxy Simulations Comparison Project \citep{2014ApJS..210...14K_short,2016ApJ...833..202K_short} we have generated a suite of isolated Milky Way-mass galaxy simulations using 9 state-of-the-art gravito-hydrodynamics codes widely used in the numerical galaxy formation community. 
In these simulations we adopted identical galactic disk initial conditions, and common physics models (e.g., radiative cooling and ultraviolet background by a standardized package). 
Subgrid physics models such as Jeans pressure floor, star formation, supernova feedback energy, and metal production were carefully constrained.
Here we release the simulation data to be freely used by the community. 
In this release we include the disk snapshots at 0 and 500 Myr of evolution per each code as used in \cite{2016ApJ...833..202K_short},  from simulations with and without star formation and feedback. 
We encourage any member of the numerical galaxy formation community to make use of these resources for their research --- for example, compare their own simulations with the {\it AGORA} galaxies, with the common analysis {\tt yt} scripts used to obtain the plots shown in our papers, also available in this release.  
\end{abstract}

\keywords{galaxies: formation -- galaxies: evolution -- methods: numerical -- hydrodynamics}


\vspace{10mm}

\section{The {\it AGORA} Initiative:  Past, Present, and Future} 

Since its launch in 2012, the {\it AGORA} High-resolution Galaxy Simulations Comparison Project ({\it Assembling Galaxies of Resolved Anatomy}) has taken aim at carefully comparing high-resolution galaxy simulations on multiple code platforms widely used in the contemporary galaxy formation research.\footnote{See the Project website at \url{http://www.AGORAsimulations.org/} for more information about the {\it AGORA} Collaboration. \label{agora-website}}  
The main goal of this initiative has been to ensure that physical assumptions are responsible for any success in galaxy formation simulations, rather than manifestations of particular numerical implementations, and by doing so, to collectively raise the predictive power of numerical galaxy formation studies. 
Over 160 individuals from over 60 different academic institutions worldwide are participating or participated in the collaborative effort of the Project.

The first result by the {\it AGORA} Collaboration \citep[][hereafter Paper I]{2014ApJS..210...14K_short} was our flagship paper which explained the philosophy behind the initiative and detailed the publicly available Project infrastructure we have assembled. 
Also described was the proof-of-concept test, in which we field-tested our infrastructure with a dark matter-only cosmological zoom-in simulation using the common initial condition \citep[generated with {\sc Music};][]{MUSIC}, finding a robust convergence amongst participating codes. 
In the second paper of the Project \citep[][hereafter Paper II]{2016ApJ...833..202K_short}, we focused on the evolution of an isolated Milky Way-mass galaxy. 
All participating codes shared the common initial condition \citep[generated with {\sc Makedisk};][]{gadget2}, common physics models \citep[e.g., radiative cooling and extragalactic ultraviolet background provided by the standardized package {\sc Grackle};][]{2017MNRAS.466.2217S}, and common analysis platform \citep[{\tt yt} toolkit;][]{yt}. 
Subgrid physics models such as Jeans pressure floor, star formation, supernova feedback energy, and metal production were carefully constrained across code platforms.
With a spatial resolution of 80 pc that resolves the scale height of the disk, we find that any intrinsic inter-code difference is small compared to the variations in input physics such as supernovae feedback. 

Through workshops and teleconferences, and with common infrastructures built together, the {\it AGORA} Collaboration has initiated a one-of-a-kind, open forum where users of different simulations codes can talk to and learn from one another.  
This platform allows the members (and non-members) to validate one another's work, promoting collaborative and reproducible research essential in any scientific community.    
Armed with the common infrastructures established, {\it AGORA} now serves as a launchpad to initiate ambitious,  astrophysically-motivated comparisons using cosmological zoom-in simulations. 
Currently we are working on the gravito-hydrodynamical simulations of a $M_{\rm vir}\sim10^{12}\,{\rm M}_{\odot}$ halo at redshift 0. 
This new suite of simulations will be fully described in an upcoming paper from the {\it AGORA} Collaboration in 2020  (Roca-F\`{a}brega et al. in prep). 
For these simulations we have adopted most of the subgrid physics and simulation strategies developed for Paper II, with improvements such as a recalibrated stellar feedback prescription and the most recent version of the {\sc Grackle} library.  
The data from these new models will be utilized to undertake a number of sub-projects based on multi-platform comparison. 
The first of the sub-projects we launched focuses on the properties and evolution of the circumgalactic medium (CGM) and its dependence on the numerical scheme.  


\section{Public Release of the Isolated Disk Simulation Data} 

Here we provide the simulation snapshots used in the analysis of Paper II.
The cohort of widely-used, state-of-the-art galaxy simulation codes who contributed to this release includes:  the Lagrangian smoothed particle hydrodynamics codes {\sc Changa} \citep[e.g.,][]{Menon15}, {\sc Gadget-3} \citep[e.g.,][]{2012MNRAS.419.1280C, 2017MNRAS.466..105A}., {\sc Gasoline} \citep[e.g.,][]{2017MNRAS.471.2357W} and {\sc Gear} \citep[e.g.,][]{revaz_computational_2016}, and the Eulerian adaptive mesh refinement codes {\sc Art-I} \citep[e.g.,][]{2014MNRAS.442.1545C}, {\sc Art-II} \citep[e.g.,][]{2013ApJ...770...25A}, {\sc Enzo} \citep[e.g.,][]{2014ApJS..211...19B} and {\sc Ramses} \citep[e.g.,][]{ramses}, and the mesh-free finite-volume Godunov code {\sc Gizmo} \citep[e.g.,][]{hopkins2015}. 
The provided snapshots are at 0 Myr (right after each code processed the initial condition) and at 500 Myr, from two different runs, the first one in which star formation is not activated and the second one with star formation and feedback.
This release will allow any interested party in the community to be able to, for example, compare their own simulation snapshots with the {\it AGORA} snapshots, using the publicly available analysis scripts on the {\tt yt} platform.  
They may also study the properties of the {\it AGORA} galaxies in coordination with the Collaboration. 

The simulation data and the common analysis scripts in {\tt yt} used to obtain the figures and diagnostics presented in Paper II are available through the Project website.\footnote{\url{http://www.AGORAsimulations.org/} or \url{http://sites.google.com/site/santacruzcomparisonproject/blogs/quicklinks/}}  
Also available in the same link are isolated and cosmological initial conditions generated by the {\it AGORA} Collaboration for galaxy simulations, and the links to the key softwares.  
We encourage all members of the numerical galaxy formation community to freely make use of these resources for their  research.

\bibliography{refs}




\end{document}